# Science Factionalism: How Group Identity Language Affects Public Engagement with Misinformation and Debunking Narratives on a Popular Q&A Platform in China

Kaiping Chen, Yepeng Jin, Anqi Shao
University of Wisconsin-Madison

**Abstract**

Misinformation and intergroup bias are two pathologies challenging informed citizenship. This paper examines how identity language is used in misinformation and debunking messages about controversial science on Chinese digital public sphere, and their impact on how the public engage with science. We collected an eight-year time series dataset of public discussion (N=6039) on one of the most controversial science issues in China (GMO) from a popular Q&A platform, Zhihu. We found that both misinformation and debunking messages use a substantial amount of group identity languages when discussing the controversial science issue, which we define as *science factionalism* – discussion about science is divided by factions that are formed upon science attitudes. We found that posts that use science factionalism receive more digital votes and comments, even among the science-savvy community in China. Science factionalism also increases the use of negativity in public discourse. We discussed the implications of how science factionalism interacts with the digital attention economy to affect public engagement with science misinformation.

Since the 2016 U.S. presidential election, research interest on mis(dis)information in political communication has grown due to concerns over its spread and influence during the election cycle (Freelon & Wells, 2020; Weeks & de Zuniga, 2021). The COVID-19 pandemic has brought a new wave of studying misinformation that cut across science and politics (Chen, Chen, Zhang, Meng, & Shen, 2020). Paralleled with the spread of misinformation is the use of group identity language, from the rise of the far-right group spreading conspiracy theories on social media (Lewis, 2018), to nationalist narratives on science issues such as labeling COVID-19 as the Kung Flu. The prevalent use of identity language from politics to science issues can be exacerbated under a high-choice media environment, where individuals selectively choose media sources and content that reinforce existing beliefs (Hameleers & van der Meer, 2020; Stroud, 2008). Group identity language and misinformation can bring severe consequences that challenge public understanding of science, obstruct informed citizenship, and fuel violence and hatred among different racial and country groups.

Despite the prevalent use of identity languages in misinformation from politics to science issues (Hart & Nisbet, 2012; Li & Su, 2020), little is understood about the consequences of identity languages on public engagement with science on social media platforms, not only in the amount of participation but also in how the public discuss about science. For instance, we have little empirical evidence examining how the public engages with identity languages in discussing misinformation. Furthermore, most research that study identity language focus on misinformation and fake news (Gertz, & Walter, 2020; Li & Su, 2020; Schulz, Wirth, & Müller, 2020), seldom do we know how identity language is used in debunking narratives, even though social media is a playground where both narratives co-exist and compete. This *comparative* understanding is critical for designing effective message campaigns to debunk misinformation. Our paper fills this gap by 1) providing a theoretical explanation on how misinformation might use identity language compared to debunking and other information, 2) explaining the consequences of group identity language in affecting how the public deliberate and engage with misinformation, and 3) collecting and analyzing a novel dataset to demonstrate the relationship between group identity language and public engagement with science for misinformation and debunking messages.

We collected a time-series dataset of public discussion on one of the most controversial science issues in China (i.e., GMO) from a popular Q&A platform, Zhihu (like its American counterpart Quora). To understand the use of group identity language in public deliberation about science and misinformation on this digital platform, we performed content analysis. Our results showed that both misinformation and debunking messages use a substantial amount of group identity languages that are based on issue attitudes. We term this phenomenon as *science factionalism* -- discussion about a science issue utilizes faction-based identity language where the publics categorize themselves into factions that support or oppose a science issue. We found that posts that use science factionalism receive more public engagement in the form of endorsement and comments, even among the science-savvy community in China. Science factionalism has consequences on public discourse, in terms of negativity and the use of justification. Diverging from most literature that found use of identity language tends to be associated with misinformation, we found that people who debunked misinformation also use a similar amount of identity language and are even more negative in discourse. By infusing group identity into thinking and discussions about science, science becomes politicized. Therefore, we argue that is it important to integrate investigations of this critical politicization vector, — group identity — into misinformation research and in designing corrective information.

Our paper is organized in the following manner. We first drew from social identity and cognition theory and literature on public engagement with science to theorize how misinformation might use group identity language and how group identity might impact public engagement with misinformation. We then presented a novel dataset we collected from a controversial science topic on a popular Q&A platform in China and introduced how we performed content analysis to understand the use of group identity language, attitude on misinformation, and the use of negativity and justification in public discourse. The article then proceeds with the major findings for our hypotheses, followed by the discussion and implication for studying misinformation and public engagement with science on social media.

## Theory and Literature Review

Social identity and misinformation are two core topics in communication and social science research. To theorize the relationship between identity and misinformation, we first drew from social identity theory to illustrate how people's identification with groups can be associated with misinformation in online public discussion. To study the consequences of group identity language in public engagement with misinformation and counterarguments, we examined social cognition studies that explicate how group identity cues within messages can be a shortcut for information processing, and thus affect how the public engage and discuss about science. We borrow from existing measurements from public deliberation and social media literature to focus on the use of negativity and justification in online discussion.

*Identity above Facts: How Group Identity is Closely Connected with Misinformation*

Group identity is a classic concept extensively studied in social psychology literature where scholars explicated how ingroup and outgroup identity can affect public opinion (Tajfel, 1981, 2010; Tajfel, 1978). These identity cues embedded in messages are likely to enhance people's social identification with individuals highlighted in the message and will in turn influence their reactions (Hart & Nisbet, 2012). There are *two* approaches to examine group identity. One common method is through surveying participants which social groups they categorize themselves into such as partisanship, race, and ethnicity. However, when it comes to studying users on social media, users often do not put their identity information in their profiles. Therefore, the other approach scholars have recently used is to examine the frequency of group identity language mentioning through content analysis of public discourse (Chinn & Hart, 2020; Hart, Chinn, & Soroka, 2020; Li & Su, 2020). In this paper, we study group identity from this *second* approach as our context of inquiry is public discussions on a social media platform where content data is available to researchers and user-level data is very limited and is thus hard to know how users categorize themselves.

There are multiple dimensions scholars have raised to study the use of group identity language in social media content. Some scholars study group identity language looking into the use of group pronouns such as "we" versus "them" (e.g., (Li & Su, 2020). Other scholars examined group name references such as partisanship (e.g., "Republicans" and "Democrats" in (Hart et al., 2020), country actors (e.g., "China" and "US" in (Chen et al., 2020), and science deniers (e.g., anti-vaxxers, climate deniers). We highlight that group identity is a multi-dimensional concept, and it is context-dependent when we want to define who the ingroup is and who the outgroup is. We will discuss which dimension we focused under the Context of Inquiry section.

To theorize the relationship between group identity and misinformation, *social identity theory* provides a useful framework, especially in explaining the ingroup and outgroup dynamics. Group identity, as defined from social cognitive studies, is formed when people position themselves in relationships with others sharing similarities (Billig & Tajfel, 1973). The social identity theory suggests a mechanism that one's positions within a group can be enhanced by ascribing negative aspects to outgroups (Tajfel & Turner, 1986). Under this mechanism, people identify with groups are inclined to make positive evaluations of ingroups and are more friendly to ingroup members (i.e., ingroup favoritism), whereas their evaluations towards outgroups tend to be hostile or negative (i.e., outgroup derogation).

This biased nature associated with group identities can provide a fertile ground for misinformation. Recent evidence suggests that group identity-related languages are found to prevail in public discussion of "fake news" (Li & Su, 2020). This "fake news" which often covers political debates, involves either endorsement for or arguments against "particular partisan identities or positions" (Marwick, 2018). A closer examination at the origin and spread of misinformation reveals that under the faith of "we are the people", individuals may generate false consensus within a group; under the faith of "you are the fake news", corrective information may be falsely regarded as misinformation to vilify the other group (Schulz, Wirth, & Müller, 2020). This prevalence of group labels has been found in debates not limited to political "fake news", but across several studies in science issues (Bruggemann, Elgesem, Bienzeisler, Gertz, & Walter, 2020; Howarth & Sharman, 2015). Group identity labels attached to people are often associated with negative implications, which hinder deliberative arguments and motivate people to put group identity or partisanship over facts in public debates (Mourao & Robertson, 2019). Therefore, in misinformation, group identity languages can be a sign of misunderstanding towards the out-groups, or false beliefs in favor of one's group affiliation.

In contrast to the flourishing works that examine group identity language in misinformation and fake news, there lacks empirical research on how group identity language exists in debunking messages. In the context of climate change, Howarth and Sherman (2015) provided an overview that unfavorable group languages are not only one-sidedly associated with misbeliefs (e.g., calling people acknowledging climate change as "alarmists"), but also widely leveraged by the opposite side (e.g., labeling people with uncertainty or skepticism as "deniers" and "contrarians"). This binary polarization in group language usage implies that attaching negative attributes to the opposite group identity is not the patent of misinformed citizens. However, we cannot assert in which condition those labels of "deniers" or "contrarians" would be used in climate change discussions. In other words, although there is ripe evidence for the prevalent use of group languages in misinformation, it is little known to what extent group identity cues are used in messages that debunk misinformation. This *comparative approach* of understanding group language use in misinformation and its counter-narratives is critical as social media is a contestation playground where conspiracy theories and debunking co-exist and compete (Jung, Ross, & Stieglitz, 2020). We, therefore, propose our first research question as:

*RQ1*. Is misinformation more likely to use group identity language, compared to debunking messages and other messages?

*How Identity Language is Associated with Public Engagement with Science*

The prevalent use of group identity cues in misinformation can have consequences on public engagement with different issues. Extensive literature suggests that individuals holding a strong identity affiliation are more likely to engage in identity-related discussions. For instance, politicized group identity can boost political engagement (Simon & Klandermans, 2001). People with strong national identities are more likely to engage in political activities and support muscular responses to national threats (Huddy, 2013). For voting behavior as a most prevalent example in research, increased engagement is found in people highly identifying themselves with certain ethnical groups (Kuo, Malhotra, & Mo, 2017; Towler, Crawford, & Bennett, 2020), gender orientations (Bankert, 2020), and partisan affiliations (Dinas, 2014).

Most literature above studies the relationship between group identity and engagement where individuals' identity can be collected through the survey methodology. However, when it comes to social media settings, often, users' identities are hard to be known. Our paper shifts the angle of *studying users' identities* to studying *the use of identity language in messages*.

We examine whether messages can receive more engagement when they contain identity cues. As most literature on social media engagement encounters, it is often difficult to know who are engaging with these messages. Therefore, majority literature examines user engagement as a whole entity rather than looking into who engages with what content (i.e., the consistency between message senders and engagers). We want to acknowledge this limitation we have as other literature does. Despite the limitation of not being able to know who engage with these identity narratives, understanding the effect of identity cues on the overall public engagement is valuable as this allows researchers to investigate how users process identity-related messages. When examining the use of identity cues in messages on the overall engagement, scholars found increased engagement with media contents (e.g., comments, recommendations, etc.,) when the media narrative uses the partisanship slant (Muddiman & Stroud, 2017; Stroud, Van Duyn, & Peacock, 2016). A recent media content study also revealed that Twitter messages containing outgroup references received more retweets from users (Rathje, Van Bavel, & van der Linden, 2021).

Theories from cognitive fields provide a key explanation to explain why people engage more frequently with messages that contain identity cues. People tend to think intuitively and attempt to reduce the complexity of decision-making processes with heuristics (Tversky & Kahneman, 1974). Group language is utilized as a cue for heuristic information processing. Political scientists attribute this phenomenon to people's *motivated reasoning* – that people would pursue a directional goal to affirm their pre-existing beliefs when faced with identity threats (Trevors, 2019). Within groups, people want to be "socially safe" and selectively expose themselves to information that reinforces their prepositions (Shu, Sliva, Wang, Tang, & Liu, 2017). Salient identity language in messages fulfill this motive by signaling this possible safety zone. Group identity featured in messages can enhance individuals' self-confirmation and reduce their inclination to carefully examine the message (Lee, 2007), which could make opinions extreme and polarized (Wojcieszak, 2010). With empirical evidence from prior research and the cognitive mechanism in mind, we raise our first hypothesis as:

*H1.* Messages using group identity language will receive more public engagement (e.g., #clicks, comments), compared to those without identity cues.

*Identity Cues and Public Discourse in Misinformation and Debunking Messages*

To capture the characteristics of public discussions, inclusiveness, civility, rationality, and interactivity are suggested as the four essential dimensions (Friess & Eilders, 2015). We will explain in this part why misinformation with identity cues is more likely to use negative language (e.g., "name-calling, contempt, and derision of the opposition" (Brooks & Geer, 2007) and less rationality (i.e., lacks the use of justification).

*Negative messages.* Social identity theory explains why identity language might also be more negative. People strengthen their position in a group by standing firmly with the prevalent belief within their group while ascribing "negative aspects" or even intergroup insults towards outgroups (Bruggemann et al., 2020; Rains, Kenski, Coe, & Harwood, 2017; Tajfel & Turner, 1986). This bias can result in negativity when the public discuss issues using group cues, such as focusing on the negative aspects of outgroups (Brooks & Geer, 2007), or negative emotional valences (e.g., (Iyengar, Sood, & Lelkes, 2012).

The association between identity cues and negativity has been extensively studied. Negative and even uncivil partisan discussion is prevalent in news comments to denounce the opponents and is considered as a form of identity performance especially in visually anonymous contexts such as online discussions (Muddiman & Stroud, 2017; Rains et al., 2017). One example of using negative messages towards outgroup members is the use of "inflammatory and superfluous" content (Brooks & Geer, 2007). In a recent study, using outgroup pronouns (e.g., "they" and "their") is shown to be more likely to collocate with hatred languages such as "contempt, blame, and hatred targeting the outgroup"(Li & Su, 2020). In the context of misinformation propagation, while misinformation itself is widely leveraged as a weapon to attack outgroups (Stockings, 2019), the accusation of "fake news" is often towards the outgroups as well (Schulz et al., 2020). Informed by the social identity theory that misinformation discussions that use group cues are also more likely to be negative, we raise our hypothesis as:

*H2.* For messages that propagate misinformation, using identity language is associated with more use of negativity.

In the context of misinformation debunking, however, there is little research on how negativity and identity language is correlated. Some studies suggested that misinformation corrections should attend to respondents' identity, otherwise misinformation correction could backfire (Nyhan & Reifler, 2010). Some recent evidence also started to explore the use of negative tones in debunking messages. They show that the effect of debunking is not dependent on whether the message is negative or not (Bode, Vraga, & Tully, 2020). Although this small group of literature noted the importance of understanding the role of identity and negative tones in designing correction messages, these experimental studies did *not* examine the prevalence of negativity and identity language in public discourse, nor did they investigate the relationship between identity languages and negativity valence in debunking narratives. Exploring this relationship is crucial as coupling negativity with identity cues could pose a challenge for respondents' worldviews (Lewandowsky, 2017), which is a premise for the backfire effect. Therefore, our paper also examines the relationship between identity language and negativity in debunking messages. Due to the very limited literature, we pose a research question here for the association between identity language and the use of negativity in misinformation debunking:

*RQ2.* For messages that debunk misinformation, is using identity language associated with more use of negativity?

*Messages lacking justification.* In addition to negativity, scholars suggested that misinformation with identity cues might also lack the use of justification. Compared with drawing hasty conclusions, public deliberation requires a justification for each choice (Burkhalter, Gastil, & Kelshaw, 2002). As aforementioned, group identity cues signaling similar positions can hinder people from examining the messages. In homogenous conversations when participants converse with people of similar group identity, people hardly justify or critically examine the information that is coherent with group beliefs, and thus lack "genuine deliberation" that is based on competing, reasoning, and exchanging arguments on different viewpoints (Strandberg, Himmelroos, & Gronlund, 2019). Thoughtful deliberation is contradictory to intergroup bias, as the former can facilitate one's accurate belief instead of further strengthening the partisan biases (Bago, Rand, & Pennycook, 2020). These findings suggest that people who use group cues for processing information would not bother making the further justification for the misleading messages.

In contrast to the identity shortcut embedded in misinformation propagation, debunking requires adequate use of justification to persuade the audience. As scholars suggested, misinformation corrections are more likely to be effective for persuasion when they provide enough justification (Lewandowsky, Ecker, & Cook, 2017), and give respondents the chance to deliberate about their opinions (Bago et al., 2020). Similar to our arguments for developing RQ2, for debunking messages in real-world public discourse such as the organic discussion on social media, it is little understood about the association between identity cues and the use of justification. Therefore, we propose a hypothesis and a research question for misinformation and debunking messages separately:

*H3.* For messages that propagate misinformation, using identity language is associated with using less justification.

*RQ3.* For messages that debunk misinformation, is using identity language associated with using more justification?

To measure the use of justification in public discourse, scholars have suggested that the speaker should validate the argument by "citing a rule of logic"(Meyers, Brashers, & Hanner, 2000). However, justification is not limited to citing scientific facts, especially the rise of social media brings more diverse knowledge types and sometimes alternative facts. Moreover, as a reason-giving process, a variety of communication styles including facts and storytelling can be used in people's opinion exchange with others (Chen, 2020; Gutmann & Thompson, 2004: 3). Narrative reasoning (i.e., storytelling) with concrete personal stories, is more favored among the public, compared with numerical reasonings with statistical information (Caple, 2019). Narrative messages in science communication have their advantage in engaging the general public, compared with scientific reasoning that can easily confuse the lay publics in the discussion (Dahlstrom & Scheufele, 2018). This study values both citing exterior information sources and personal narratives as means of justification.

### *Context of inquiry: Identity Language in GMO Discourse in China*

To study the use of identity language in discussing misinformation, this paper focuses on the Chinese social media context and we study misinformation related to one of the most controversial science issues in China -- GMO (Liang, Liu, & Zhang, 2019). GMO discussion on social media has sustained in the past ten years. The GMO discussion landscape on Chinese social media differs from that in U.S. and European countries since the controversy in GMO under the Chinese context is entangled with its political, cultural, and social contexts. For instance, many GMO misinformation in China is highly related to the cultural belief of Daoism that pursues "naturalness" and little human intervention. Consequently, many criticisms toward GMO center around its genetic engineering aspect. The divergent attitudes and opinions generated towards GMO gradually evolved into factions: the pro-GMO faction and the anti-GMO faction, among Chinese netizens (Gao, 2016; Yan, 2013).

Furthermore, GMO discussion in China is also shaped by issue entrepreneurs that represent the scientist community *vs* the journalist community (Chen & Jin, 2021), where public discussion reached a climax after a debate between a famous journalist Cui and a university professor in genetics Lu at a public lecture event (China Science Net, 2015, March). Therefore, among many public discussions about this lecture, we observe that the publics categorize themselves into factions: Fans of Professor Lu (belonging to the pro-GMO faction) *vs* Fans of Journalist Cui (belonging to the anti-GMO faction)[1]. Because of the prevalent reference to faction language about this science issue, this paper takes a focus of studying identity language through examining the reference to the pro-GMO and anti-GMO faction on Zhihu's discourse. As we discussed in the literature section, identity language is a multi-dimensional concept, and thus, we also acknowledge the value of studying GMO discourse using other identity dimensions in our Discussion section.

## Data and Method

*Data Collection*

Among different online platforms for discussing and spreading science information and misinformation, Zhihu is the one with a unique social media structure where group identity plays a critical role in technology discussion (Hamm & Lin, 2019; Peng et al., 2009). Similar to its US counterpart Quora, Zhihu allows users to contribute knowledge to online communities by asking and answering questions, commenting, and voting for each other's answers (Jin, Li, Zhong, & Zhai, 2015; Wang & Shah, 2016).

Zhihu is one ideal platform to examine how the public engage and discuss about science topics. First, the votes can serve as a perfect proxy to measure the popularity of the answers. The recommendation algorithm encourages users to generate popular (i.e., highly-voted) answers to attract more followers. We are curious about whether this popularity can signal higher discourse quality. Secondly, commonly perceived as a high-end knowledge-sharing community, Zhihu is widely praised for its rigorous and evidence-based writing styles in the top voted answers in science topics (Liang et al., 2019; Zhang, 2018). Zhihu has no limitation of answer length, thus users can cite external information sources to justify their arguments. This characteristic not only provides us with variations of justification source, but also utilizes a systematical examination of the attitudes and behaviors among better-educated and "rationale think" groups. Third, similar to the US Q&A platform Quora (Wang, Gill, Mohanlal, Zheng, & Zhao, 2013), Zhihu is the place where the distribution of answers and questions is highly skewed. A small amount of people would answer a major number of questions. Thus, we limit our research scope to the answers created by the most active users and track how active users varied their discourse content on social media.

We used the Zhihu Web Crawler API[2] to collect all the answers about GM (转基因), which result in 40,101 answers posted by more than 21,422 unique users as of September 2019. Then we chose the users that have posted more than 10 GMO-related answers and got 221 unique authors who have written 6039 answers in total for in-depth content analysis. The Zhihu API allows us to obtain the author id, posted date, and the number of votes and comments an answer received. To measure public engagement on social media, we borrow from social media studies that examine user metrics such as the number of views, votes, likes, and comments (Bode, 2017; Skoric, Zhu, Goh, & Pang, 2016; Xu, Yu, & Song, 2018). Our operationalization of the public engagement metrics is limited by the data that the platform provides to researchers.

---

[1] For some example discussions where netizens categorized GMO groups into factions, see: https://www.zhihu.com/question/301071836; https://www.douban.com/group/topic/73682351/?author=1； http://blog.sciencenet.cn/blog-362400-711082.html

[2] https://zhihu-oauth.readthedocs.io/zh_CN/latest/

On the Zhihu platform, users can vote for or against an answer. However, the API only gives researchers the data about "vote for", and thus the number of votes is more like public endorsement (i.e., click "like")[3]. Zhihu API also gives researchers the number of comments each answer received. User behaviors like clicks and commenting are all metrics to measure various levels of engagement (Aldous, An, & Jansen, 2019) and in this paper, we examined all the user metrics data that is available to us (including #of vote for and comments).

*Content Analysis*

Figure 1 summarizes our content analysis variables and procedures. For these 6039 answers, we first manually coded them along two dummy variables: Mention Group and Mention Misinformation. *Mention Group* means that an answer mentions the anti/pro-GMO group (e.g., using language like "anti-GMO faction" (反转派，崔粉) or "pro-GMO faction" (挺转派，卢粉). These phrases are commonly used on social media to refer to the different factions that support or oppose GMO in China. We manually annotated all our answers to examine whether it mentions group or not.

To determine whether GMO misinformation is mentioned in each answer, we referred to common misinformation and corresponding science facts about GMO summarized by the Chinese Agricultural Biosafety Management Office, the China Association for Science and Technology (China MOA., 2014), and other literature (China CDC., 2015; Huang, Qiu, Bai, & Pray, 2006; Liang & Xing, 2015; Yan, 2012). *Mention Misinformation* means that an answer discussed any of the four types of GMO-related misinformation: (1) the false health risk of GM food (Huang et al., 2006; Scott, Inbar, Wirz, Brossard, & Rozin, 2018), (2) the fake environment effects from GMO plantation (MOA., 2017), (3) people's misperception of government regulation of GMO (Yang, Xu, & Rodriguez, 2014), and (4) GMO-related international political misinformation (Skeptical Explorer, 2018).

If an answer was coded as *Mentions GMO-related Misinformation*, we further coded users' *Attitudes* on the Misinformation and their *Use of Justifications* in these misinformation related answers. Different from other social media platforms where many laypeople spread misinformation about GMO and the minority is debunking misinformation, Zhihu is known for its highly educated netizens or at least science-savvy users who engage with debunking unreliable misinformation. Many answers that *mention GMO misinformation* correct the misinformation while some propagate the misinformation. Thus, we coded users' attitudes toward the misinformation along 1) Propagate Misinformation, 2) Debunk Misinformation, 3) Ambiguous Attitude toward Misinformation. Answers that showing agreement with misinformation are annotated as "propagate misinformation"; answers showing disagreement with misinformation are annotated as "debunk misinformation"; and answers showing neither agree nor disagreement are annotated as "ambiguous attitude toward misinformation".

For coding the *Use of Justification* on these misinformation related answers, we first coded whether an answer used *justification* or not. If so, we further coded the types of justification the author used: Personal Story or Exterior Information Source[4]. Two researchers conducted

---

[3] We are not able to distinguish the Pro-GMO votes vs the Anti-GMO votes because we are not able to collect the user information who cast the vote. The API gives us the total number of votes. Moreover, as we emphasized in our literature section, for this paper, we study group identity looking into the identity cues used on the post-level, rather than on the user-level, as user information on Zhihu is very limited for scholars to categorize users into different identity groups.

[4] Here we are interested at whether an answer cited external information source (e.g., mainstream media, journals/scientists,

this manual content analysis and the inter-coder agreement for each variable is high, ranging from 0.812 to 1 using Cohen's kappa coefficient. For details of our content analysis codebook and examples, see Supplemental Material Appendix I.

**Figure 1. Content analysis variables**

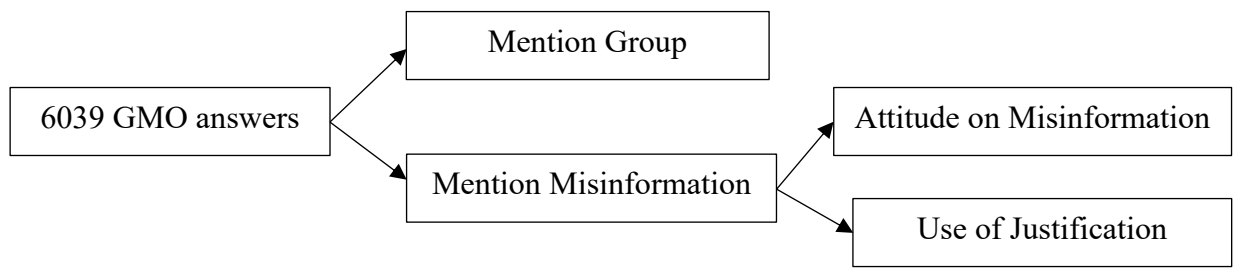

*Applying Dictionary Method to Analyze Negativity of Public Discourse*

With the rapid development of social network and text analysis, an increasing number of organizations are making use of sentiment dictionary to extract sentiment information from massive social media texts (Zhang, Wei, Wang, & Liao, 2018). We used the Chinese Affective Lexicon Ontology Dictionary developed by Xu et al., to calculate the percentage of anger and disgust emotion of Zhihu answers (Xu, Lin, Pan, Ren, & Chen, 2008). Based on Ekman's work on basic emotions (Ekman, 1992), Xu et al., (2008) classified the emotion into seven categories: Anger, Disgust, Sadness, Fear, Surprise, Happy, and Praise. As we are interested at the use of negative languages on social media, we wrote a python script to calculate the sentiment score from the "Anger" and the "Disgust" categories. Before applying the dictionary method, we also conducted human validation on a random sample of messages (van Atteveldt, van der Velden, & Boukes, 2021) by manually checking whether these answers contain the expression of anger or disgust. The accuracy of "Anger" and "Disgust" is 0.96 and 0.83, respectively.

## Result

*The Use of Group Identity Language in Misinformation Propagation and Debunking*

There are two interesting findings for RQ1 which examines the use of group identity language comparing answers that propagated misinformation with answers that debunked misinformation or did not talk about misinformation at all. First, we found that answers that discussed misinformation-related things about GMO used a much higher percentage of group identity language compared to answers that were not misinformation relevant. For instance, Table 1 shows that for those answers that did not discuss misinformation about GMO, only 7.6% used group identity language while 92.4% did not use group identity language. In

---

documentaries, social media etc.) rather than simply expressing opinions. We want to caution here that we did not verify the truthfulness of the justification because justification such as using personal storytelling is hard to verify, although scholars raised storytelling as an important element for public deliberation (Chen, 2021). For example, if the author cited an article which is proven to be inconsistent with science facts, we still code the answer as "Citing External Information Source." Future scholars can develop further content analysis variable to study the "sources" cited in a post.

contrast, we see that for the columns on misinformation-related answers (i.e., propagate, debunk, ambiguous attitude), the use of group identity language is much higher (e.g., 25.1% in those answers that propagated misinformation). This difference is not only substantive but also significant. The risk ratios indicate that when an answer propagated or debunked misinformation, there is an over three-fold increase in mentioning the group identity language compared to answers that did not talk about misinformation.

We also compared the mentioning of the Pro-GMO faction and Anti-GMO faction across different information types[5]. We found that a high proportion of Propagating Misinformation answers mentioned the Pro-GMO faction (22.7%), while a comparable high proportion of Debunking Misinformation answers mentioned the Anti-GMO Faction (22.9%). This finding provides a more nuanced look into how answers who differed in their attitude toward GMO misinformation mentioned factions differently. For answers that propagated GMO misinformation, as these users might be more likely to be anti-GMOs, they often referred to their out-group --the pro-GMO group when they propagated GMO misinformation and vice versa for answers that debunked misinformation.

Table 1. Comparing the Use of Group Identity Language across Information Types

|  | Full Sample | Misinformation-irrelevant Answers | Propagate Misinformation | Debunk Misinformation | Ambiguous Attitude toward Misinformation |
|---|---|---|---|---|---|
| N | 5858 | 4049 | 374 | 992 | 443 |
| Mention Group | 662 (11.7%) | 288 (7.6%) | 92 (25.1%) | 235 (23.7%) | 47 (10.6%) |
| Mention Pro-GMO | 233 (3.9%) | 97 (2.4%) | 85 (22.7%) | 23 (2.3%) | 28 (6.3%) |
| Mention Anti-GMO | 491 (8.4%) | 222 (5.5%) | 15 (4.0%) | 227 (22.9%) | 27 (6.1%) |
| Not Mention Group | 5196 (88.3%) | 3761 (92.4%) | 282 (74.9%) | 757 (76.3%) | 396 (89.4%) |
| Risk Ratios |  | To test the use of group identity language (Mention Group) on different subsets, we calculated the relative risk ratio of Mention Group α and its 95% confidence interval (Rothman, 2012) of the following information types (we set the Misinformation-irrelevant Answers as the baseline): Propagate Misinformation: α = 3.45, CI = [2.80,4.26] Ambiguous Attitude α = 1.49, CI = [1.11,1.99] Debunk Misinformation: α = 3.33, CI = [2.84,3.90] | | | |

*Note:* Number of observations went down from 6039 answers to 5858 as some answers did not have the public engagement indicator # of comments available to researchers. To interpret this table, we need to look at the *columns* as the percentages are calculated for the column. Taking "Propagate misinformation" for example, there are 374 answers that propagated misinformation, among them 92 answers (i.e., 92/374=25%) mentioned group identity language and 282 answers (i.e., 282/374=75%) did not mention group identity language.

*The Consequences of Group Identity Language on Public Engagement with Science*

---
[5] There could be overlapping as some answers could mention Anti-GMO and Pro-GMO group at the same time.

We examined public engagement with science from users' endorsement of an answer (i.e., # of votes an answer obtained) and the use of negativity and justification in an answer. Our findings support H1. Table 2 shows the Ordinary Least Square (OLS) regression regarding the association between use of identity language and the number of votes and comments. In the baseline model (All Answers), we found that using group identity language in answers is associated with more votes and comments, compared to those answers that did not use identity language ($\beta_{vote} = 0.62^{***}$, $\beta_{comment} = 0.73^{***}$). We repeated the regression model for different information types (e.g., misinformation irrelevant, answers that propagate misinformation, debunk misinformation, and hold ambiguous attitudes toward misinformation). We found that mentioning misinformation in an answer is always associated with a higher number of votes and comments no matter the information types.

Table 2. Group Identity Language and Public Engagement across Information Types: OLS Regression

|  | Public Engagement (Vote, Comment) | | | | | | | | | |
|---|---|---|---|---|---|---|---|---|---|---|
|  | All Answers | | Misinformation Irrelevant | | Propagate Misinformation | | Debunk Misinformation | | Ambiguous Attitudes | |
|  | (vote) | (comment) | (vote) | (comment) | (vote) | (comment) | (vote) | (comment) | (vote) | (comment) |
| Mention Group | 0.616*** | 0.729*** | 0.500*** | 0.587*** | 0.541*** | 0.958*** | 0.543*** | 0.463*** | 0.584** | 0.944*** |
| Follower | 0.164*** | 0.063*** | 0.140*** | 0.050*** | 0.222*** | 0.155*** | 0.181*** | 0.086*** | 0.162*** | 0.107*** |
| Following | 0.014 | 0.032*** | 0.031*** | 0.029*** | -0.053 | -0.120** | 0.065** | 0.127*** | 0.049 | 0.022 |
| Author Badge | 0.181** | 0.048 | 0.155** | 0.069 | 0.435 | -0.181 | 0.034 | -0.004 | 0.571** | -0.206 |
| Text Length | 0.0005*** | 0.001*** | 0.002*** | 0.002*** | 0.0001 | 0.0002** | 0.0004*** | 0.0004*** | 0.0005*** | 0.001*** |
| Year | 0.031** | -0.017 | 0.019 | -0.022* | 0.012 | -0.032 | 0.042 | -0.036 | 0.012 | 0.004 |
| Constant | -62.184** | 35.374 | -37.596 | 43.686* | -24.172 | 65.451 | -84.309 | 73.277 | -23.783 | -8.514 |
| Observations | 5,307 | 5,307 | 3,656 | 3,656 | 308 | 308 | 964 | 964 | 379 | 379 |
| R² | 0.159 | 0.118 | 0.167 | 0.127 | 0.260 | 0.133 | 0.174 | 0.129 | 0.190 | 0.161 |
| Adjusted R² | 0.158 | 0.117 | 0.165 | 0.126 | 0.245 | 0.116 | 0.169 | 0.124 | 0.177 | 0.148 |

*Note:* *p**p***p<0.01

Vote and comment represent the log (number of vote) and log (number of comments) respectively.

H2 (RQ2) and H3(RQ3) investigate the association between the use of group identity language and the use of negativity and justification, for answers that propagated misinformation and for those that debunked misinformation. For instance, an example answer that used disgust words is: "Arguments from some public intellectuals are *so awful and disgusting*. They never used their brain to think when they made these arguments…." Here is an answer that used anger to blame the anti-GMO faction: "…The reason @username stressed that all scientists supported the golden rice is because he wanted to smear the motivation of scientists. When the Anti-GMO faction's bad intention was exposed, *they were so angry and yelled on* social media (just like @username). The whole anti-GMO faction are birds of the same feather…" For the use of negative languages in public discourse, our finding does not support H2. As shown in Table 3, for answers that propagated misinformation, there is no significant association between use of group language and the use of negativity. Unexpectedly, for answers that debunked misinformation, use of group language is also associated with more use of negativity (RQ2).

Table 3. Group Identity Language and Negativity of Public Discourse across Information Types:

OLS Regression

| | Negativity of Public Discourse (Number of Anger or Disgust Words) | | | | | | | | | |
|---|---|---|---|---|---|---|---|---|---|---|
| | All Answers | | Misinformation-Related | | Propagate Misinformation | | Debunk Misinformation | | Ambiguous Attitudes | |
| | (anger) | (disgust) | (anger) | (disgust) | (anger) | (disgust) | (anger) | (disgust) | (anger) | (disgust) |
| Mention Group | 0.011 | 0.335*** | 0.027 | 0.244*** | -0.080 | -0.195 | 0.052* | 0.311*** | 0.035 | 0.307** |
| Follower | -0.0003 | 0.024*** | 0.001 | 0.027** | 0.008 | -0.019 | -0.0004 | 0.052*** | 0.006 | -0.002 |
| Following | -0.003 | -0.042*** | -0.0005 | -0.072*** | -0.030* | -0.022 | 0.006 | -0.116*** | -0.001 | 0.006 |
| Author Badge | 0.003 | -0.089** | 0.022 | -0.193* | -0.051 | 0.018 | 0.012 | -0.243* | 0.056 | -0.088 |
| Text Length | 0.0002*** | 0.0003*** | 0.0001*** | 0.0002*** | 0.0001*** | 0.0002*** | 0.0001*** | 0.0001** | 0.0002*** | 0.0003*** |
| Year | -0.003 | 0.011 | -0.008 | -0.010 | 0.026 | -0.048 | -0.014** | -0.006 | -0.008 | -0.020 |
| Constant | 5.214 | -20.706 | 15.975 | 20.906 | -52.286 | 97.779 | 29.207** | 13.277 | 16.580 | 41.074 |
| Observations | 5,307 | 5,307 | 1,651 | 1,651 | 308 | 308 | 964 | 964 | 379 | 379 |
| R² | 0.094 | 0.086 | 0.073 | 0.061 | 0.064 | 0.062 | 0.075 | 0.087 | 0.255 | 0.074 |
| Adjusted R² | 0.093 | 0.085 | 0.070 | 0.057 | 0.046 | 0.044 | 0.069 | 0.081 | 0.243 | 0.059 |

Note: *p**p***p<0.01

For the use of justification in public discourse, Table 4 compares the use of justification between answers that used group identity language and those that did not use group identity language. We found that among all the answers that mentioned group identity, 55.2% used justification, while among the answers that did not mention group identity, only 36.5% used justification. The prevalence risk ratio suggests among the Mention Group answers, the proportion of using justification is 50% higher compared to those Not Mention Group answers (α=1.51, CI= [1.35,1.69)). We further examined the relationship between the use of group identity language and use of justification for debunking answers and misinformation propagation answers separately (Table 5). We found that for answers that propagated misinformation, there is no significant association between use of group identity language and the use of justification, thus rejecting H3. For answers that debunked misinformation, use of group identity language is associated with more use of justification (RQ3).

Table 4. Comparing the Use of Justification between Answers that Use Group Identity Language versus Those that Did Not Use Group Identity Language

| | All | No Justification | Use Justification | | |
|---|---|---|---|---|---|
| | | | All | Cite Exterior Sources | Cite Personal Stories |
| All answers that discussed about GMO misinformation | N=1809 | 1077 (59.5%) | 732 (40.5%) | 711 (39.3%) | 50 (2.8%) |
| Mention Group | n=374 | 167 (44.7%) | 207 (55.3%) | 203 (54.3%) | 12 (3.2%) |
| Not Mention Group | n=1435 | 910 (63.4%) | 525 (36.5%) | 508 (35.4%) | 38 (2.6%) |
| Prevalence Risk Ratio | Prevalence risk ratio α of using Justification among Mention Group answers compared to Not Mention Group answers: α=1.51, CI= (1.35,1.69). | | | | |

Table 5. Group Identity Language and Use of Justification across Information Types: Probit Regression

|  | Justification | | | | | |
|---|---|---|---|---|---|---|
|  | All | Misinformation-Irrelevant | Misinformation-Related | Propagate Misinformation | Debunk Misinformation | Ambiguous Attitudes |
|  | (1) | (2) | (3) | (4) | (5) | (6) |
| Mention Group | 0.528*** | -3.661 | 0.334*** | 0.269 | 0.279*** | 0.011 |
| Follower | -0.004 | 0.147 | -0.008 | -0.016 | -0.011 | -0.127*** |
| Following | -0.085*** | -1.574 | -0.059*** | 0.138** | -0.109*** | 0.023 |
| Author Badge | -0.202* | -0.620 | -0.187 | -0.368 | -0.157 | -0.335 |
| Text Length | 0.001*** | 0.001 | 0.001*** | 0.0002 | 0.001*** | 0.002*** |
| Year | 0.076*** | 0.268 | 0.073*** | -0.039 | 0.093*** | 0.022 |
| Constant | -154.072*** | -544.467 | -146.609*** | 78.610 | -186.668*** | -44.374 |
| Observations | 5,307 | 3,656 | 1,651 | 308 | 964 | 379 |
| Log Likelihood | -1,550.262 | -5.001 | -975.659 | -192.961 | -566.155 | -143.450 |
| Akaike Inf. Crit. | 3,114.524 | 24.002 | 1,965.317 | 399.921 | 1,146.309 | 300.899 |

*Note:* *p**p***p<0.01

*Breaking Down the Consequences of Ingroup and Outgroup Language Use on Public Engagement with Science*

In the previous section, we showed the consequences of mentioning group identity language on public engagement and discussion of GMO misinformation. This section expands by decomposing the Mention Group variable into two sub-dimensions: mention ingroup and mention outgroup. Ingroup refers to those answers that mentioned a specific faction (anti or pro-GMO) and expressed support for the faction it mentioned. Outgroup refers to those answers that mentioned a specific faction (anti or pro-GMO) but expressed opposition toward the faction it mentioned (for details, see Supplemental Material Appendix II).

We conducted regression analyses for the ingroup and the outgroup variable separately for each DV, with the same control variables we used in the previous section. Coefficients from each regression equation are presented in Figure 2 (for full regression tables, see Supplemental Material Appendix II). The three column labels represent the information type (e.g., answers that propagate, debunk, or shows ambiguous attitudes toward GMO misinformation). The three row labels represent our dependent variables. Each of the 9 panels is separated into two parts: the left part presents the regression coefficient of our ingroup variable, and the right part presents the regression coefficient of our outgroup variable.

We want to highlight two main interesting findings. First, for the impact of in-group and out-group mentioning on public engagement (top panel in Figure 2), we found that when an answer propagated or debunked misinformation, mentioning outgroups increased the number of votes and comments an answer received. This suggests that expressing opposition toward a faction could drive engagement. Second, for answers that debunked misinformation (the middle panel in Figure 2), we found again that mentioning outgroups (i.e., opposing a faction) increased not only public engagement, but also the use of negativity. This implies that users who debunked misinformation in fact used more negative especially inflammatory languages such as anger and disgust when they opposed a faction.

**Figure 2.** The Impact of Ingroup and Outgroup Language on Public Engagement, Use of Negativity and Justification in Public Discourse

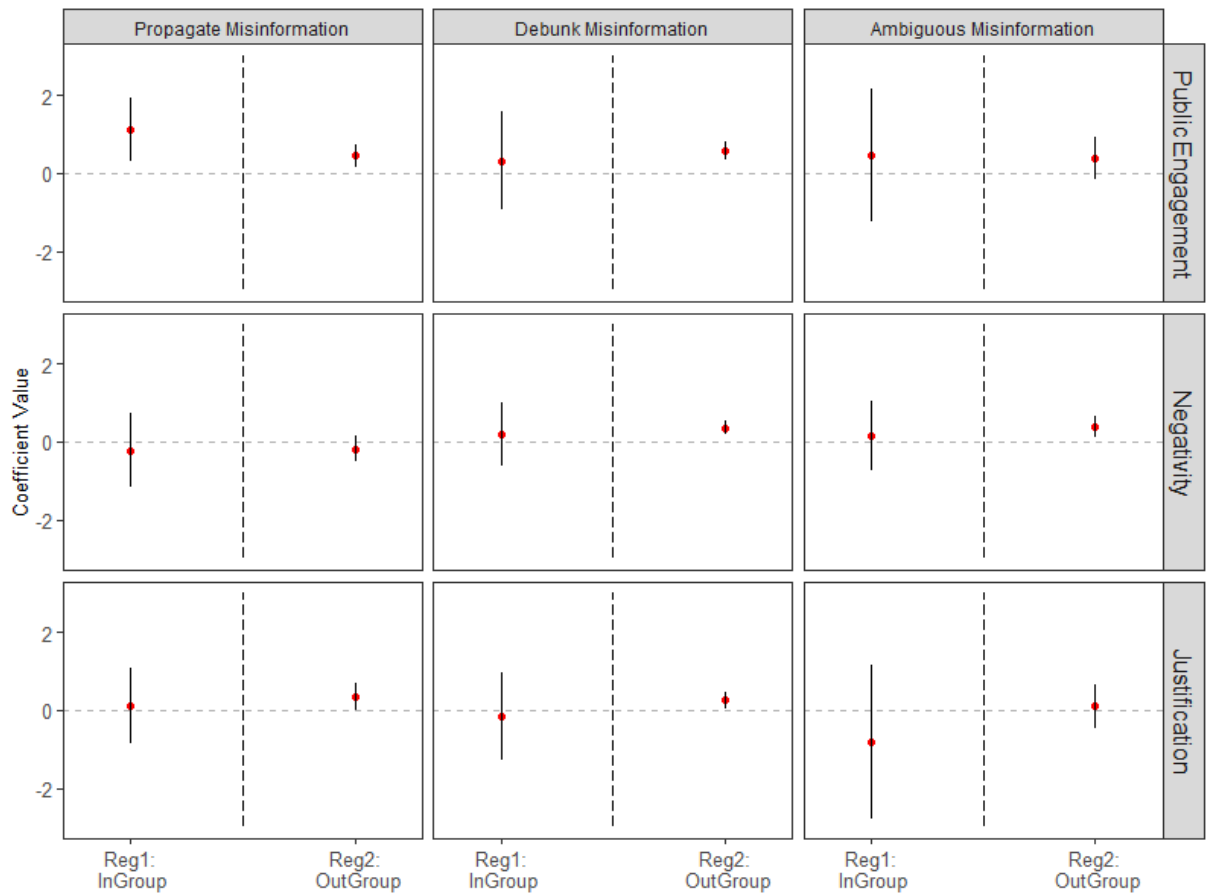

## Discussion and Conclusion

Despite the prevalent use of identity languages in misinformation from politics to science issues (Hart & Nisbet, 2012; Li & Su, 2020), little is understood about the consequences of identity languages on public engagement with science on social media platforms, not only in participation but also in how the public deliberated about the science topic. Our paper fills this gap by demonstrating how misinformation and debunking messages utilize group identity cues and the impact of group languages on public engagement with science. To do this, we collected and analyzed a rich time-series dataset about online discourse on a controversial science topic, GMO, with a regional focus on China. Our results showed that both misinformation and counter-narratives used a substantial amount of group identity languages about this controversial science issue, which we term the phenomenon as *science factionalism* – discussion about science is divided by factions that are formed based on science attitudes. We found that posts that called for science factionalism received more digital votes, even among the science-savvy community in China. Science factionalism also brings consequences to public discourse, in terms of negativity and the use of justification, depending on whether the discourse propagates or debunks the misinformation.

*Science factionalism* echoes with but also differs from literature on polarization of science (Rekker, 2021). Echoing works on polarization of science, we also identified that discussions about GMO consisted of many languages that refer to ingroups and outgroups. Science factionalism differs from polarization in several manners. It captures science communication in political systems where multi-party competition lacks or more broadly to issues that are not simply polarized by "partisanship" or "political ideology", but more by "faction", and group identity in general. This use of factional language to distinguish oneself from others is not based on partisanship or political ideology, but rather on science attitudes or attitudes toward certain stakeholders in science issues. As the social identity theory argues, identity is a multi-dimensional concept (Roccas, Sagiv, Schwartz, Halevy, & Eidelson, 2008). Compared to partisanship, faction is less stable and allows researchers to study the formation of science attitudes beyond a single identity. Thus, science factionalism allows researchers to examine science communication in countries where multi-party competitions lack, and the publics are more homogenous in political ideology.

Our findings also provide a new angle to study identity language and narratives. As highlighted in the literature review, social and group identity is a multi-dimensional concept that encompasses many aspects, from the use of pronouns (Li & Su, 2020), to partisan language (Sedona Chinn, Hart, & Soroka, 2020), to hatred language among race and ethnicity groups (Wojcieszak, 2010), and to nationalist and populist narratives in public discussions (Castanho Silva, Vegetti, & Littvay, 2017; Guan & Yang, 2020; Woods & Dickson, 2017). In this paper, we propose a new angle of studying group identity which is *domain-driven*, "science factionalism". For instance, this paper's issue focus is GMO, and the identity typology people use to refer to each other is faction: pro-GMO faction *vs* anti-GMO faction. We deliberately use the word "factionalism" instead of "nationalism" or "populism" because users on the Zhihu platform are Chinese, yet, based on their attitudinal differences on GMO, they categorize themselves into different factions. Faction is the identity language the public raises and uses and thus we want to capture this identity dimension. This domain-driven identity language in the form of public attitude divide is crucial for studying controversial science, as documented in many works where scholars use attitude to refer to certain groups in science discussions, such as climate deniers, anti-vaxxers (O'Neill & Boykoff, 2010; Weinberg & Dawson, 2020). Nevertheless, we acknowledge that it will be fruitful for future research to examine how other dimensions of identity languages (e.g., gender, rural vs urban) are played out in science discussion.

We also acknowledge that our context of inquiry is China, and our conclusions are limited to this context. Our study examines the manifest identity cues in GMO discussions. It will be valuable for future works to examine to what extent science factionalism play out in other country contexts, as some scholar has noted the use of factional languages such as GMO camp or GMO lobby in the European context (Marris, 2001).

Third, different from literature that often assumes that misinformation would use more group conflict languages (Li & Su, 2020; Schulz et al., 2020), we found that debunking narratives also used a similar amount of group reference languages on social media. This *comparative approach* toward studying science misinformation and its counter-narratives is crucial as social media provides a playground where both narratives prevail and compete. This comparative finding also poses interesting questions for future research such as why debunking narratives have been found to fail in persuading the other side (Kreps & Kriner, 2020; Nisbet & Scheufele, 2009). Maybe one potential reason is their overplay of group differences rather than seeking mutual understanding among groups.

This science factionalism is elevated by the economic model of digital media, where attention equals revenue (Myllylahti, 2020). Echoing with literature on information diffusion which found that rumors diffuse further and deeper with audience (Vosoughi, Roy, & Aral, 2018), we also found that discourse that use group references received more votes on this Q&A platform. The logic of attention economy of social media might encourage people to reinforce the use of group languages, as the more one can position him/herself in a group, the more likely his/her posts will be liked or viewed. This attention economy provides interesting implications for communicating correction messages. To reach more engagement, correction messages need to mention group identity languages. The key to avoid the failure is to mention group identity language in the correct framing. For instance, a potential strategy for debunking misinformation is to first acknowledge the common grounds of different groups before trying to correct misperception. Exploring the strategies of how to mention the differences and common grounds across groups opens an exciting agenda for future science communication research and experimentations.

There are consequences of this interaction between science factionalism with the digital economy model. As we demonstrated in this paper, answers using group cues are more negative compared to other answers, where they used words related to disgust to downgrade other groups. This negative consequence of science factionalism is expected. However, what is concerning is that Zhihu is a platform where the users are much more science-savvy compared to the lay publics (Liang et al., 2019). Yet, we observed this factionalism even among this science-savvy community. This factionalism among the science-savvy community could bring spillover effects where other lay publics might mimic these negative behavioral and discourse patterns (Chen et al., 2020).